\documentstyle[psfig]{article}
\begin{document}
\huge
\begin{center}
Arithmetic of plane Cremona transformations and the\\ dimensions of 
transfinite heterotic string space-time
\end{center}
\vspace*{.0cm}
\Large
\begin{center}
Metod Saniga

\vspace*{.3cm}
\small
{\it Astronomical Institute, Slovak Academy of Sciences, 
SK-059 60 Tatransk\' a Lomnica, Slovak Republic}
\end{center}

\vspace*{-.4cm}
\noindent
\hrulefill

\vspace*{.2cm}
\small
\noindent
{\bf Abstract}

It is shown that the two sequences of characteristic dimensions of transfinite
heterotic  string  space-time  found  by  El  Naschie  can  be remarkably well 
accounted for in terms of  the arithmetic of self-conjugate homaloidal nets of 
plane algebraic curves of orders 3 to 20. A firm algebraic geometrical 
justification is thus given not only for all the relevant dimensions of the 
classical theory, but also for other two dimensions proposed by El Naschie, 
viz. the inverse of quantum gravity coupling constant ($\simeq 42.36067977$) 
and that of (one half of) fine structure constant ($\simeq 68.54101967$). 
A non-trivial coupling between the two El Naschie sequences is also revealed.

\noindent
\hrulefill

\vspace*{.4cm}
\large
\noindent
{\bf 1. Introduction}

In  a  recent  series  of  papers  [1--9], El Naschie has demonstrated that the
transfinitely-extended heterotic string space-time exhibits two unique 
sequences of fractal dimensions
\begin{equation}
D^{{\rm A}}_{q}=\bar{\alpha}_{0} \phi^{q}
\end{equation}
and
\begin{equation}
D^{{\rm B}}_{q}=\frac{\bar{\alpha}_{0}}{2} \phi^{q},
\end{equation}
where  $q$ is a positive integer, $\bar{\alpha}_{0}$  is the inverse value of 
the fine structure constant and $\phi$ represents the Hausdorff dimension of 
the elementary (kernel) Cantor set. Taking  $\bar{\alpha}_{0}=136+6\phi^{3} 
\left(1- \phi^{3} \right)$  and  $\phi$ to be identical with the golden mean,
$\phi= 1 - \phi^{2} = \left(\sqrt{5} - 1 \right)/2 \simeq  0.618033989$,  Eqs. 
(1) and (2) yield [2,5,8,9]

\vspace*{.2cm}
\begin{center}
\begin{tabular}{ccccccccc}
\hline \hline 
$q$ & (0) & 1 & 2 & 3 & 4 & 5 & 6 & 7 \\ 
\hline 
$D^{{\rm A}}_{q}$ & $(136+6k)$ & $84+4k$ & $52+2k$ & $32+2k$ & ~20~ & 
$12+2k$ & $8-2k$ & $4+4k$  \\
$D^{{\rm B}}_{q}$ & $(68+3k)$ & $42+2k$ & $26+k$ & $16+k$ & ~10~ & $6+k$ 
& $4-k$ & $2+2k$ \\ 
\hline \hline 
\end{tabular}
\end{center}
\vspace*{.2cm}

\noindent
where $k=\phi^{3}\left(1-\phi^{3}\right) \simeq 0.18033989.$ For $k=0$ the 
second sequence is formally recognized to encompass all the relevant 
dimensions of the {\it classical} heterotic string theory [10]. Soon after we 
had become familiar with El Naschie's work, we noticed that the sequence in 
question bears, for $q=2,3,\ldots, 6$, an extraordinary close resemblance to 
the sequence of the number of lines lying on (ordinary) Del Pezzo surfaces 
[11,12]. Motivated by this observation, we raised a question whether there 
exists an algebro-geometrical structure that would, at least partially, 
reproduce {\it both} the above-introduced sequences. And such a structure was, 
indeed, found. It is associated with the concept of plane Cremona 
transformations and, as we will see in what follows, it offers a nice fit to 
both El Naschie's sets simultaneously.

\vspace*{.6cm}
\noindent
{\bf 2. Plane Cremona transformations and self-conjugate homaloidal nets}

A plane Cremona transformation [13] is a birational correspondence between 
the points of two 
projective planes $P_{2}$ and $P_{2}'$, being generated, in either plane, by a 
specific family of algebraic curves. This family possesses three characteristic 
properties [13,14]: 1) it is linear and doubly-infinite (the so-called net);
2) any two distinct curves of it have one and only one free intersection, i.e.
the intersection which is different from any base (i.e. shared by all the 
members of the family) point; and
3) all the curves are rational, i.e. birationally transformable into 
lines.
A net of curves meeting these three constraints is called 
{\it homaloidal}. For curves of any given order $n$, these homaloidal 
nets are uniquely characterized by the total number and multiplicities of their base 
points. The order of a Cremona transformation is the order $n$ of the curves
of its generating homaloidal net ${\cal N}$. It can easily be verified that if
${\cal N}$ and ${\cal N'}$ are the homaloidal nets generating, respectively,
a given Cremona transformation ($P_{2} \rightarrow P_{2}'$) and its inverse
($P_{2}' \rightarrow P_{2}$), then they must be of the same order. 
Furthermore, the total number of base points is also the same in each plane.
Yet, the two nets ${\cal N}$ and ${\cal N'}$ are, in general, {\it not} of
the same nature. If they differ from each other, they are called conjugate;
if they are identical, they are called {\it self}-conjugate. And it is the
latter that serve our purpose here.

\vspace*{.6cm}
\noindent
{\bf 3. Arithmetic of self-conjugate nets and El Naschie's sequences}

Our next attention is exclusively focussed on how the total number of 
self-conjugate homaloidal nets of a given order $n$, $\#_{n}^{{\rm sc}}$,
depends on the value of $n$. For the first 21 orders this relation was
found as early as 1922 by B. Mlodziejowski [15], and in a tabular form it
looks as follows:

\vspace*{.2cm}
\begin{center}
\begin{tabular}{cccccccccccccccccccccc}
\hline \hline 
$n$ & 1 & 2 & 3 & 4 & 5 & 6 & 7 & 8 & 9 & 10 & 11 & 12 & 13 & 14 & 15 & 16
& 17 & 18 & 19 & 20 & 21  \\ 
\hline 
$\#_{n}^{{\rm sc}}$ & 1 & 1 & 1 & 2 & 3 & 2 & 3 & 5 & 4 & 5 & 7 & 9 & 10 & 11 
& 11 & 14 & 18 & 16 & 21 & 32 & 27  \\
\hline \hline 
\end{tabular}
\end{center}
\vspace*{.2cm}

\noindent
At first sight, there seems to be nothing particular about the progression
$\#_{n}^{{\rm sc}}$ on its own. It becomes attractive for us only after we 
define the quantities
\begin{equation}
D_{l}^{(2)} \equiv \#_{3l+1}^{{\rm sc}} + \#_{3l+2}^{{\rm sc}}
\end{equation}
and
\begin{equation}
D_{l}^{(3)} \equiv 
\#_{3l}^{{\rm sc}} + \#_{3l+1}^{{\rm sc}} + \#_{3l+2}^{{\rm sc}},
\end{equation}
with $l$ being a positive integer, which, for the first six values
of $l$, are found to acquire the following interesting values:

\vspace*{.2cm}
\begin{center}
\begin{tabular}{lclcl}
\hline \hline 
$l$ & $D_{l}^{(2)}$ & ($D_{q}^{{\rm A}}$) & $D_{l}^{(3)}$  & 
($D_{q}^{{\rm B}}$)  \\ 
\hline 
1 & 5  & (\hspace*{.19cm}4.721\ldots) & 6  & (\hspace*{.19cm}6.180\ldots) \\
2 & 8  & (\hspace*{.19cm}7.639\ldots) & 10 & (10.000\ldots) \\
3 & 12 & (12.360\ldots) & 16 & (16.180\ldots)\\
4 & 21 & (20.000\ldots) & 30 & (26.180\ldots)\\
5 & 32 & (32.360\ldots) & 43 & (42.360\ldots)\\
6 & 53 & (52.360\ldots) & 69 & (68.541\ldots)\\
\hline \hline 
\end{tabular}
\end{center}
\vspace*{.2cm}

\noindent
We  see  that  $D_{l}^{(2)}$ is an amazingly good integer-valued match of the 
$D_{q}^{{\rm A}}$  sequence  for  $2 \leq q \leq 7$  (compare  the second and
third column of the table),  whilst  $D_{l}^{(3)}$  does the same job for the  
$D_{q}^{{\rm B}}$  series,  here  for $0 \leq q \leq 5$ (compare the last two
columns  of  the  table).  In  both  the cases, the only relevant discrepancy 
between our ``homaloidal" sequences and those of El Naschie occurs for $l=4$: 
whereas  in the former case this difference is rather slight ($21 - 20 = 1$), 
in  the  latter  case  it  is  more  pronounced ($30 - (26 + k) = 4- k$). But 
especially the latter fact should not disturb us at all, for
\begin{equation}
D_{l=4}^{(3)} = 30 = 26 + 4 = 26 + k + 4 - k = D_{q=2}^{{\rm B}} +
D_{q=6}^{{\rm B}},
\end{equation}
that is, $D_{l=4}^{(3)}$,  instead  of representing a {\it single} dimension, 
can equivalently be looked upon as  the (exact) sum of {\it two}, and perhaps 
most important,  dimensions  of  (both  classical  and transfinite) heterotic 
string space-time!
\footnote{This  is  perhaps most intriguing and unexpected observation in the 
present  paper,  which  may well turn out to provide us with invaluable clues
for a further development of the theory of heterotic string space-time in its
transfinite generalization.}

There  are  several  implications  of  serious  physical  importance that the
above-described findings impart on us. The first one is the fact that both El 
Naschie's  hierarchies  of  fractal  dimensions should be treated on the {\it 
same}  par,  i.e.  both  should  be  regarded  as  {\it equally}  relevant in 
describing the structure of transfinite heterotic string  space-time. This is 
simply  a  result of the  common  origin of both the ``homaloidal" sequences,
embodied in the arithmetic of the set $\#_{n}^{{\rm sc}}$, $3 \leq n \leq 20$.
The second fact is related to the value of $D_{l=5}^{(3)}=43$. This dimension 
is obviously  an  integer-valued match of  $D_{q=1}^{{\rm B}} = 42 + k \simeq
42.36067977$, which is regarded by  El  Naschie [9]  to be identical with the 
smallest  possible  value of the inverse of quantum gravity coupling constant 
$\bar{\alpha}_{{\rm  g}}$  in  the  non-supersymmetric  case;  our  second 
``homaloidal" sequence thus gives a first explicit, algebraic geometrical {\it 
justification}  of  the  relevance  of $\bar{\alpha}_{{\rm g}}$ for heterotic 
strings,  as  envisaged  and  kept  enthusiastically emphasized by El Naschie 
[1--9]. Third, it should not go unnoticed that $D_{l=6}^{(3)}=69$ gives equal 
relevance to the last, and highest, dimension in the series, viz. $D_{q=0}^{{
\rm B}} = 68 + 3k \simeq 68.54101967$.  As the double of the latter value is 
thought to be very close to the inverse value of the fine structure constant 
[2], our findings also {\it justify} the prominent role this constant is 
supposed to play, as already recognized by El Naschie [16], in all fundamental 
theories  of  stringy  space-times  and Cantorian space ${\cal E}^{(\infty)}$ 
as well. The final outcome of our analysis that deserves to be properly 
underlined is the {\it coupling} between the two El Naschie fractal sequences. 
Here it is important to realize that $D_{q}^{{\rm B}}$ is not linked physically 
with $D_{q}^{{\rm A}}$, as Eqs. (1) and (2) would seem to imply, but -- as it 
can easily be discerned from comparison of the last table with the first one 
-- rather to $D_{q+2}^{{\rm A}}$. So, $D_{q+2}^{{\rm A}}$ and $D_{q}^{{\rm 
B}}$ can be viewed as twin/paired dimensions. Out of these, we are then 
naturally led to form new sequences, the simplest ones being, of course, those 
obtained by subtracting and adding the counterparts. While the first operation 
brings up nothing new, as $D_{q}^{\ominus} \equiv D_{q}^{{\rm B}} - D_{q+2}^{{
\rm A}} = D_{q+3}^{{\rm B}}$, the other one, $D_{q}^{\oplus}\equiv D_{q}^{{\rm 
B}}+D_{q+2}^{{\rm A}}$, is a new sequence: its most relevant terms look like

\vspace*{.2cm}
\begin{center}
\begin{tabular}{cllllllll}
\hline \hline 
$q$ & 1 & 2 & 3 & 4 & 5 & 6 & 7 \\ 
\hline 
$D^{\oplus}_{q}$ &  $74+4k$ & $46+k$ & $28+3k$ & $18-2k$ & 
$10+5k$ & $8-7k$ & $2+12k$  \\ 
= & 74.721\ldots & 46.180\ldots & 28.541\ldots &
 17.639\ldots &  10.901\ldots &  6.737\ldots &  4.164\ldots \\
\hline \hline 
\end{tabular}
\end{center}
\vspace*{.2cm}

\noindent
and, interestingly, they are found to copy very closely the recently
discovered by C. Castro (hierarchy of) fractal dimensions generated by 
transfinite {\bf M}-theory [17].

\vspace*{.6cm}
\noindent
{\bf 4. Summarizing conclusion}

Employing the algebra and arithmetic of self-conjugate homaloidal nets of 
planar algebraic curves, we have discovered a couple of integer-valued 
progressions that are found to mimic extraordinary well the two El Naschie 
sequences of fractal dimensions characterizing transfinite heterotic string 
space-time. One of the progressions is demonstrated to provide an algebraic 
geometrical justification not only for all the relevant dimensions of the 
classical theory, but also for other two dimensions advocated by El Naschie, 
namely the smallest value of the inverse of non-supersymmetric quantum gravity 
coupling constant, $\bar{\alpha}_{{\rm  g}} = 42+k \simeq 42.36067977$, and 
the inverse of fine structure constant, $\bar{\alpha}_{0}=2 \otimes (68+3k) 
\simeq 2 \otimes 68.54101967$. In addition, our ``homaloidal" arguments  
elucidate how the two El Naschie sequences are coupled to each other and 
imply the existence of a third ``fundamental" fractal sequence, which bears 
an intriguing similarity to the hierarchy of fractal dimensions emerging from 
a transfinite extension of the {\bf M}-theory.

\vspace*{.6cm}
\noindent
{\bf Acknowledgements}

I am extremely indebted to Mrs. Bella Shirman, a senior librarian of the
University of California at Berkeley, for her kind assistance in tracing and 
making for me a copy of Ref. 15. This work was supported in part by the NATO 
Collaborative Linkage Grant PST.CLG.976850.

\end{document}